\input{aipcheck}
\documentclass[
    ,draft            
    ,numberedheadings 
  ]
  {aipproc}
\layoutstyle{6x9}
\newcommand{\BR}{\mathcal{B}}

\newcommand{\ppbar}{p \overline{p}}

\newcommand{\pin}{\pi^-}

\newcommand{\etap}{\eta^{\prime}}
\newcommand{\g}{\gamma}
\newcommand{\ar}{\rightarrow}

\newcommand{\psp}{\psi^{\prime}}
\newcommand{\pspto}{\psi^{\prime}\to}
\newcommand{\jpsi}{J/\psi}
\newcommand{\jpsito}{J/\psi\to}
\newcommand{\pspp}{\psi^{\prime\prime}}
\newcommand{\psppto}{\psi^{\prime\prime}\to}
\newcommand{\chicz}{\chi_{c0}}

\newcommand{\chicJ}{\chi_{cJ}}
\newcommand{\ppkk}{\pi^+\pi^-K^+K^-}
\newcommand{\EE}{e^+e^-}

\newcommand{\pip}{\pi^+}
\newcommand{\pim}{\pi^-}
\newcommand{\piz}{\pi^0}
\newcommand{\ppp}{\pi^+\pi^-\pi^0}
\newcommand{\kap}{K^+}
\newcommand{\kam}{K^-}
\newcommand{\ks}{K^0_s}

\newcommand{\rhopi}{\rho\pi}

\begin{document}

\title{Hadron Spectroscopy from BES and CLEOc}



\author{Chang-Zheng YUAN}{
  address={Institute of High Energy Physics,
  Chinese Academy of Sciences,
  Beijing, China}
}

\begin{abstract}

Recent results from BES and CLEOc experiments on hadron
spectroscopy and charmonium decays using $\jpsi$, $\psp$ and
$\pspp$ data samples collected in $\EE$ annihilation are reviewed,
including the observation of $X(1835)$ in $\jpsi\to \gamma \pip
\pim \etap$, study of the scalar particles in $\jpsi$ radiative
and hadronic decays, as well as in $\chicz$ hadronic decays, and
the study of the ``$\rho\pi$ puzzle'' in $\jpsi$, $\psp$, and
$\pspp$ decays.

\end{abstract}

\maketitle


\section{Introduction}

BESII~\cite{bes} running at BEPC and CLEOc~\cite{cleo} running at
CESR are the two detectors operating in the $\tau$-charm energy
region, and have collected large data samples of charmonium decays
including 58~M $\jpsi$ events, 14~M $\psp$ events, and
33~pb$^{-1}$ data around $\pspp$ peak at BESII, and 4~M $\psp$
events, and 281~pb$^{-1}$ $\pspp$ events at CLEOc. To study the
continuum background in the charmonium decays, special data
samples at center of mass energy lower than the $\psp$ mass were
taken both at BESII ($\sqrt{s}=3.65$~GeV) and at CLEOc
($\sqrt{s}=3.671$~GeV), the luminosity are 6.4~pb$^{-1}$ and
21~pb$^{-1}$ respectively. These data samples are used for the
study of the hadron spectroscopy, the $D$ decay properties and the
CKM matrix, as well as the charmonium decay dynamics.

In this paper, we focus on the search for the new resonances in
$\jpsi$ decays, the properties of the scalars from $\jpsi$
radiative and hadronic decays, and a new approach of studying the
scalars using $\chicJ$ decays, and the extensive study of the
``$\rhopi$ puzzle'' related physics in $\jpsi$, $\psp$ and $\pspp$
decays.

It should be noted that the CLEOc detector is much better than the
BESII detector, especially in the photon detection, this makes its
4~M $\psp$ events data sample produces results with similar
precision as from 14~M $\psp$ events from BESII.

\section{Observation of $X(1835)$}

\begin{figure}[hbtp]
    \includegraphics[height=.3\textheight]{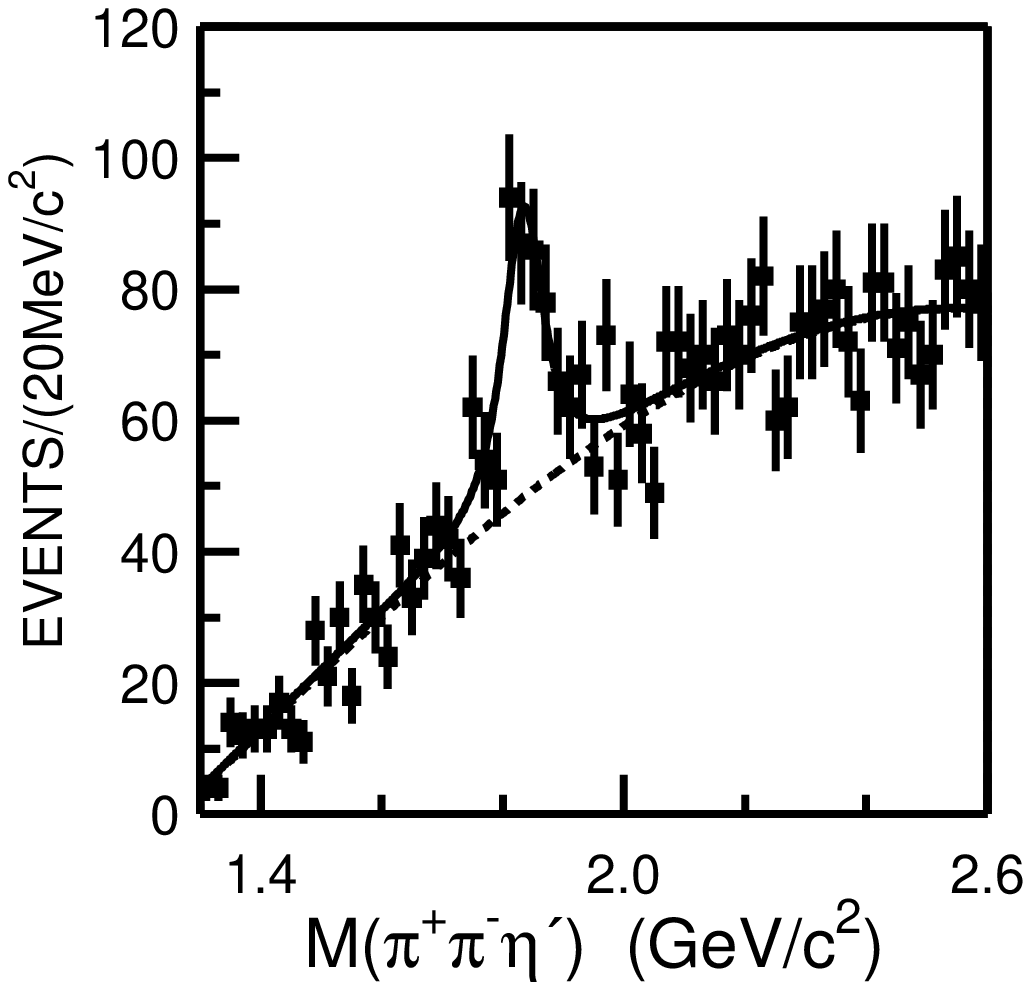}
          \caption{The $\pip\pin\etap$ invariant mass distribution for
          selected $\jpsi\ar\g\pip\pin\etap(\etap\ar\pip\pin\eta,\eta\ar\g\g)$
          and $\jpsi\ar\g\pip\pin\etap(\etap\ar\g\rho)$ events.
          Dots with error bars are data, solid line is the fit,
          and the dashed curve indicates the background function. }
    \label{sum}
    \end{figure}

The decay channel $\jpsi\to \gamma \pip \pim \etap$, with
$\etap\ar\pip\pin\eta$, or $\etap\ar\g\rho$, is analyzed using a
sample of $58\times 10^6$ $\jpsi$ events collected with the BESII
detector~\cite{etap}, to search for the other decay modes of the
possible existing $p\bar{p}$ bound state as observed in
$\jpsi\ar\g p \bar{p}$ process at BESII~\cite{gpp}.
Figure~\ref{sum} shows the $\pip\pin\etap$ invariant mass spectrum
for the combined $\jpsi\ar\g\pip\pin\etap(\etap\ar\pip\pin\eta)$
and $\jpsi\ar\g\pip\pin\etap(\etap\ar\g\rho)$ samples. A clear
peak is observed at around 1.835~GeV/c$^2$. The spectrum is fitted
with a Breit-Wigner (BW) function convolved with a Gaussian mass
resolution function (with $\sigma = 13$~MeV/c$^2$) to represent
the $X(1835)$ signal plus a smooth polynomial background function.
The signal yield from the fit is $264\pm 54$ events and the
statistical significance is 7.7~$\sigma$.

The mass of $X(1835)$ is measured to be $M = 1833.7\pm
6.1(stat)\pm 2.7(syst)$~MeV/c$^2$, the width is $\Gamma = 67.7\pm
20.3(stat)\pm 7.7(syst)$~MeV/c$^2$, and the product branching
fraction is $B(\jpsi\ar \g X)\cdot B(X\ar \pip\pin\etap)$ =
$(2.2\pm 0.4(stat)\pm 0.4(syst)) \times 10^{-4}$. The mass and
width of the $X(1835)$ are not compatible with any known meson
resonance listed by PDG~\cite{pdg}. In Ref.~\cite{pdg}, the
candidate closest in mass to the $X(1835)$ is the (unconfirmed)
$2^{-+}$ $\eta_2(1870)$ with $M=1842\pm 8$~MeV/c$^2$. However, the
width of $225\pm 14$~MeV/c$^2$, is considerably larger than that
of the $X(1835)$ (see also~\cite{BESI}, where the $2^{-+}$
component in the $\eta\pi\pi$ mode of $\jpsi$ radiative decay has
a mass $1840 \pm 15$~MeV/c$^2$ and a width $170\pm 40$~MeV/c$^2$).
Another candidate with mass close to the $X(1835)$ in
PDG~\cite{pdg} is the $X(1860)$ observed in the $p\bar{p}$ mass
threshold in radiative $\jpsi\ar\g p \bar{p}$ decays~\cite{gpp},
where a mass of $1859^{+6}_{-27}$~MeV/c$^2$, and width smaller
than 30~MeV/c$^2$ at 90\% C.L. were reported. It has been pointed
out that the $S$-wave BW function used for the fit in
Ref.~\cite{gpp} should be modified to include the effect of
final-state-interactions (FSI) on the shape of the $p\bar{p}$ mass
spectrum ~\cite{fsi1, fsi2}. By redoing the $S$-wave BW fit to the
$p\bar{p}$ invariant mass spectrum of Ref.~\cite{gpp} including
the Isospin zero, $S$-wave FSI factor of Ref.~\cite{fsi2}, BES
gets a mass $M = 1831 \pm 7$~MeV/c$^2$ (in good agreement with the
$X(1835)$) and a width $\Gamma < 153$~MeV/c$^2$ at the 90$\%$ C.L.
(not in contradiction with the $X(1835)$). There are also
theoretical arguments that the two states, $X(1835)$ and
$X(1860)$, are indeed one particle, and is a $\ppbar$ bound
state~\cite{yan,zhu}, however, other possible interpretations of
the $X(1835)$ that have no relation to the $\ppbar$ mass threshold
enhancement are not excluded. Further information about the
$X(1835)$ and $X(1860)$, especially better precisions on the mass
and width measurements and the determination of the spin-parity
are strongly desired before concluding the nature of the states.

\section{Scalar particles in $\jpsi$ and $\chicz$ decays}

The study of the scalars are very important in two aspects: in
experiment, there are still controversies about the resonance
parameters of these states; and in theory, it is still hard to
incorporate all the experimental results in a self-consistent
picture. The reason for the former is somewhat due to the
techniques used in extracting the physics information from the
experimental data, namely, the partial wave analysis (PWA) was
extensively used in the analyses, but sometimes it is rather
arbitrary what resonance need to be included in the complicated
fit with a few ten and even more than one hundred free parameters.
The reason for the latter, in part is due to the fact that the
experimental results may not all be reliable, and the complexity
in the low energy QCD domain that the higher order terms neglected
may not be small, and the mixing of the states in principle is
hard to be considered completely.

\subsection{Radiative and hadronic $\jpsi$ decays}

\begin{figure}
  \includegraphics[height=0.65\textheight]{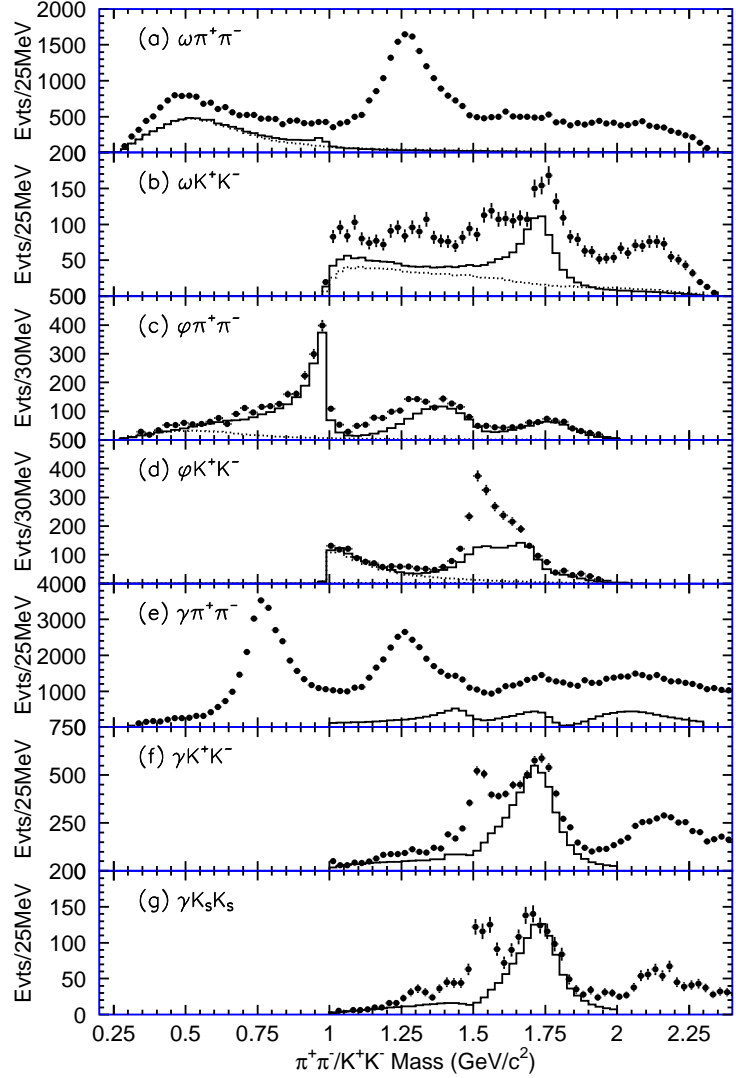}
  \caption{The invariant mass distributions of the pseudoscalar meson
pairs recoiling against $\omega$, $\phi$, or $\gamma$ in $\jpsi$
decays measured at BESII. The dots with error bars are data, the
solid histograms are the scalar contribution from PWA, and the
dashed lines in (a) through (c) are contributions of $\sigma$ from
the fits, while the dashed line in (d) is the $f_0(980)$. Notice
that not the full mass spectra are analyzed in (e), (f), and (g).
Results in (e) are preliminary, otherwise are published.}
\label{scalars}
\end{figure}

Using the world largest $\jpsi$ data sample in $\EE$ annihilation
experiment, BES studied the scalars decay into pair of
pseudoscalars ($\pip\pim$, $\piz\piz$, $\kap\kam$ and $\ks\ks$) in
$\jpsi$ radiative decays as well as recoiling against a $\phi$ or
an $\omega$~\cite{wpipi,wkk,phipipi,gkk}. The full mass spectra
and the scalar part in them are shown in Fig.~\ref{scalars}.

From the analyses, BES sees significant contributions of $\sigma$
particle in $\omega\pip\pim$ and $\omega\kap\kam$, and also hint
in $\phi\pip\pim$. Two independent partial wave analyses are
performed on $\omega\pip\pim$ data and four different
parameterizations of the $\sigma$ amplitude are tried, all give
consistent results for the $\sigma$ pole, which is at  $(541 \pm
39) - i(252 \pm 42)$~MeV/c$^2$. There is also events accumulation
in the low $\pip\pim$ mass in $\gamma \pip\pim$ mode, most
probably due to the contribution of the $\sigma$, however, there
is no attempt to analyze the structure at BES, one possible reason
is the presentation of the huge background from $\rho^0\piz$.
Nevertheless, the coupling of the $\sigma$ with a photon in
$\jpsi$ decays is an important piece of information for the
understanding of the nature of the particle, a better detector
with more statistics may want to measure it. A measurement of the
resonance parameters of $\sigma$ using the BESII $\psp\to \jpsi
\pip\pim$ data gives similar pole position as measured in
$\jpsi\to \omega \pip\pim$~\cite{zhuys}.

Strong $f_0(980)$ is seen in $\phi \pi ^+\pi ^-$ and $\phi K^+K^-$
modes, from which the resonance parameters are measured to be $M =
965 \pm 8(stat) \pm 6(syst) $~MeV/c$^2$, $g_1 = 165 \pm 10(stat)
\pm 15(syst)$~MeV/c$^2$ and $g_2/g_1 = 4.21 \pm 0.25(stat) \pm
0.21(syst)$, where $M$ is the mass, and $g_1$ and $g_2$ are the
couplings to $\pi\pi$ and $K\bar{K}$ respectively if the
$f_0(980)$ is parameterized using the the Flatt\'e's formula. The
production of $f_0(980)$ is very weak recoiling against an
$\omega$ or a photon, which indicates $s\bar{s}$ is the dominant
component in it.

The $\phi \pip \pim$ data also show a strong scalar contribution
at around 1.4~GeV/c$^2$, it is due to the dominant $f_0(1370)$
interfering with a smaller $f_0(1500)$ component. The mass and
width of $f_0(1370)$ are determined to be: $M = 1350 \pm
50$~MeV/c$^2$ and $\Gamma = 265 \pm 40$~MeV/c$^2$. In $\gamma \pip
\pim$, a similar structure is observed in the same mass region,
the fit yields a resonance at mass $1466\pm 6(stat)\pm
16(syst)$~MeV/c$^2$ with width of $108^{+14}_{-11}(stat) \pm
21(syst)$~MeV/c$^2$, possibly the $f_0(1500)$, and the
contribution from the $f_0(1370)$ can not be excluded. The
production of $f_0(1370)$ and $f_0(1500)$ in $\gamma K \bar{K}$ is
insignificant.

The $K^+K^-$ invariant mass distributions from $\gamma K \bar{K}$
and $\omega K^+K^-$, the $\pip\pim$ invariant mass distributions
from $\gamma \pip\pim$, and $\phi\pip\pim$ show clear scalar
contribution around 1.75~GeV/c$^2$. Two states are resolved from
the bump, one is $f_0(1710)$ with $M \sim 1740$~MeV/c$^2$ and
$\Gamma\sim 150$~MeV/c$^2$ which decays to $K \bar{K}$ mostly, and
one possible new state $f_0(1790)$ with $M \sim 1790$~MeV/c$^2$
and $\Gamma\sim 270$~MeV/c$^2$ which couples to $\pi\pi$ stronger
than to $K\bar{K}$. However, the existence of the second scalar
particle needs confirmation: the signal observed in $\phi
f_0(1790)$ is rather in the edge of the phase space, and the
reconstruction efficiency of the $\phi$ decreases dramatically as
the momentum of the $\phi$ decreases thus the momentum of the kaon
from $\phi$ decays is very low and can not be detected.
Furthermore, there are wide higher mass scalar states above
2~GeV/c$^2$ as observed in $\gamma \pip\pim$ (Fig.~\ref{scalars}e)
and $\gamma K \bar{K}$~\cite{pdg}, whose tails may interfere with
the $f_0(1710)$ and produce structure near the edge of the phase
space.

The use of these measurements for understanding the nature of the
scalar particles can be found in
Refs.~\cite{oset,bugg,close,zhao}, where the $\jpsi$ decay
dynamics and the fractions of the possible $q\bar{q}$ and glueball
components in the states are examined.

\subsection{Pair production of scalars in $\chicz\to \ppkk$}

Partial wave analysis of $\chi_{c0} \rightarrow \pi^+ \pi^- K^+
K^-$ is performed~\cite{ppkk} using $\chicz$ produced in $\psp$
decays at BESII. In 14~M produced $\psp$ events, 1371
$\psi^{\prime} \rightarrow \gamma \chi_{c0}$, $\chi_{c0}
\rightarrow \pi^+\pi^-K^+K^-$ candidates are selected with around
3\% background contamination.

\begin{figure}[b]
\begin{minipage}{7cm}
 \resizebox{15pc}{!}{\includegraphics[height=.5\textheight]{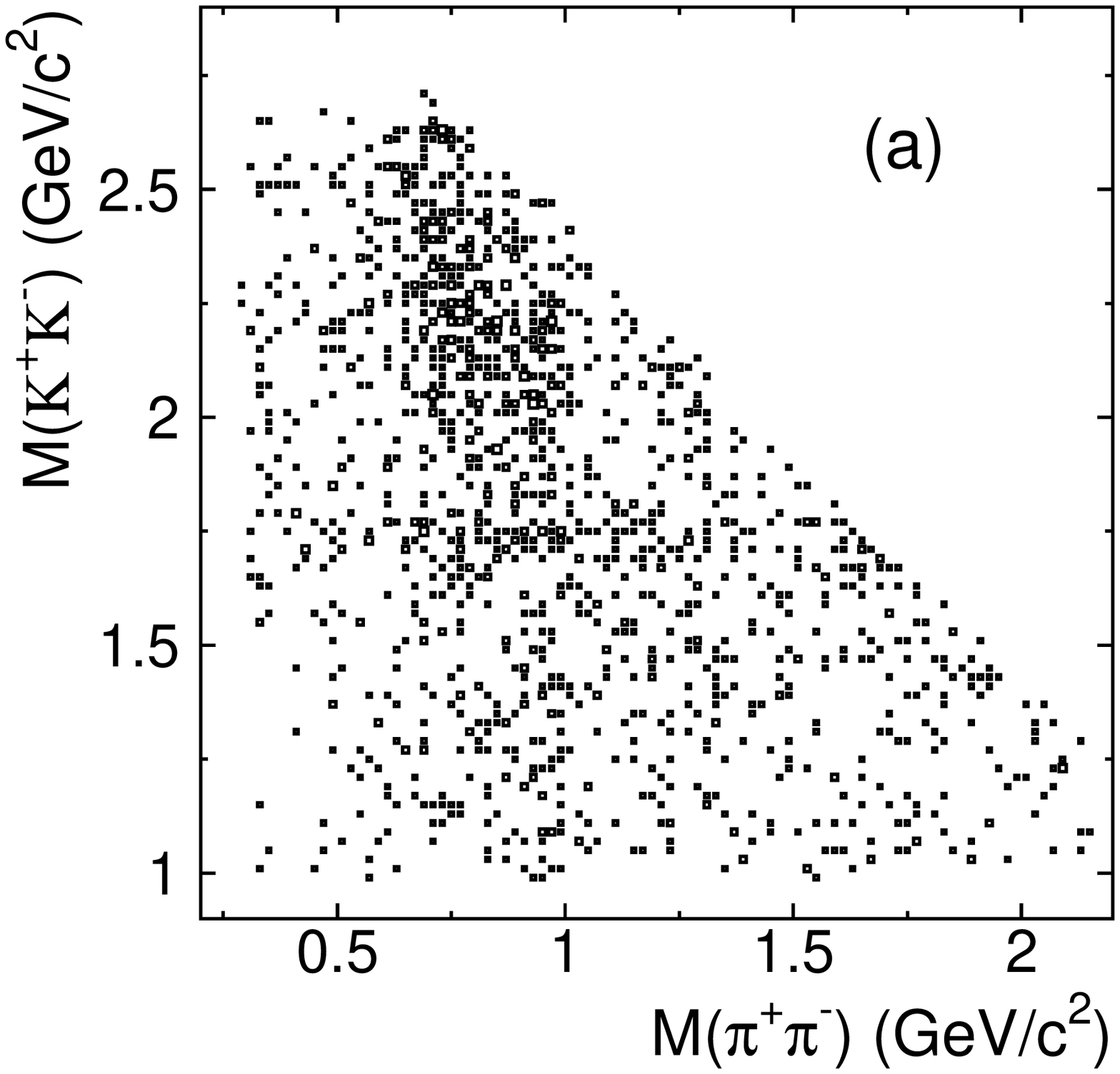}}
\end{minipage}
\begin{minipage}{7cm}
 \resizebox{15pc}{!}{\includegraphics[height=.5\textheight]{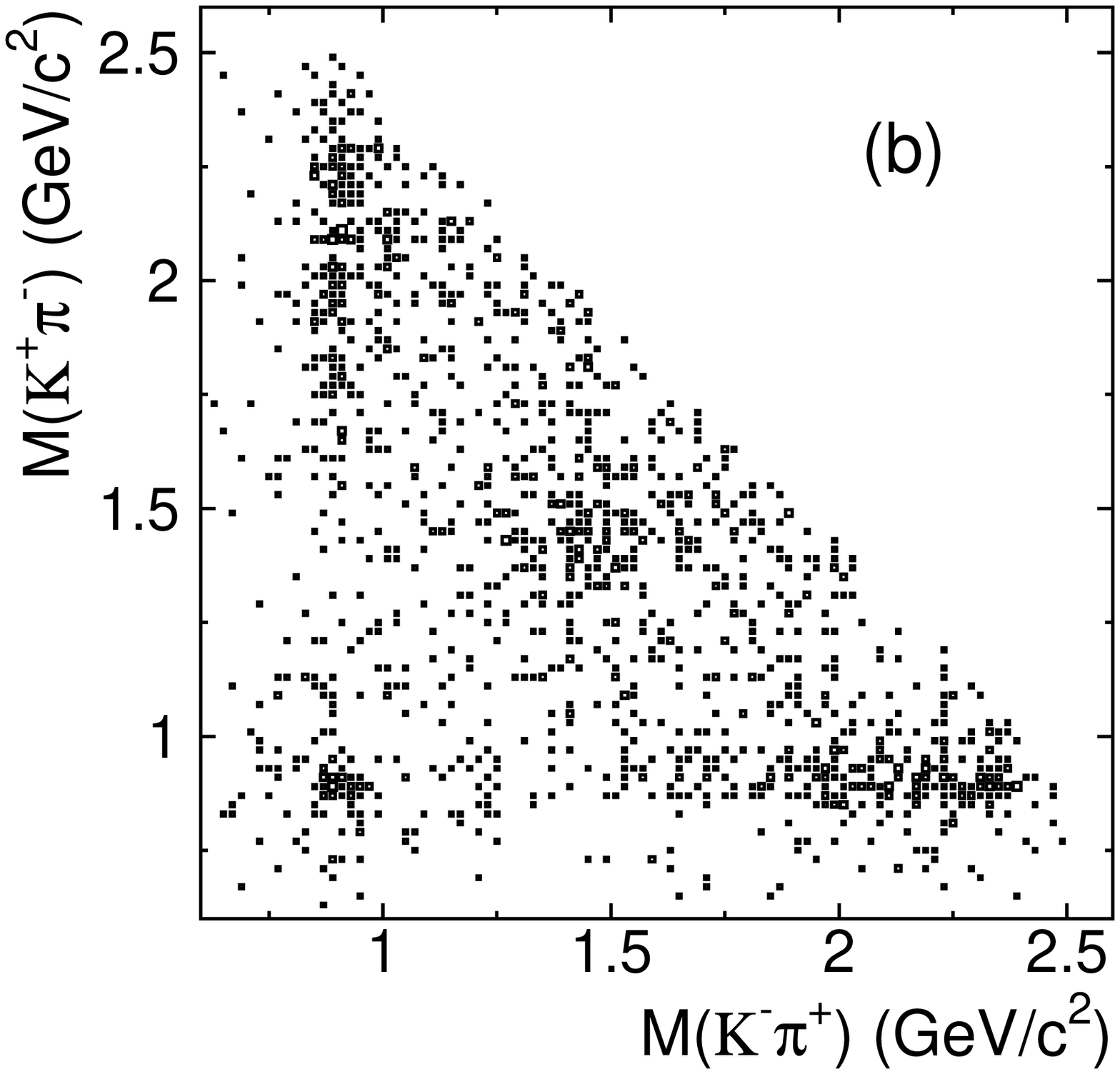}}
\end{minipage}
\caption{\label{fig_xchi0} The scatter plots of (a) $K^+K^-$
versus $\pi^+\pi^-$ and (b) $K^+\pi^-$ versus $K^-\pi^+$ invariant
mass for selected $\psi^{\prime} \rightarrow \gamma\chi_{c0}$,
$\chi_{c0} \rightarrow \pi^+\pi^-K^+K^-$ events.}
\end{figure}

Fig.~\ref{fig_xchi0}(a) shows the scatter plot of $K^+K^-$ versus
$\pi^+\pi^-$ invariant mass which provides further information on
the intermediate resonant decay modes for $(\pi^+\pi^-)(K^+K^-)$
decay, while Fig.~\ref{fig_xchi0}(b) shows the scatter plot of
$K^+\pi^-$ versus $K^-\pi^+$ invariant masses for the resonances
with strange quark.

Besides $(\pi\pi)(KK)$ and $(K\pi)(K\pi)$ modes, $(K\pi\pi)K$ mode
which leads to a measurement of $K_1(1270)K$ and $K_1(1400)K$
decay processes is also included in the fit. The PWA results are
summarized in Table~\ref{pwachi}. From these results, we notice
that scalar resonances have larger decay fractions compared to
those of tensors, and such decays provide a relatively clean
laboratory to study the properties of scalars, such as $f_0(980)$,
$f_0(1370)$, $f_0(1710)$, and so forth. The upper limits of the
pair production of the scalar mesons which are less significant
are determined at the 90\% C.L. to be {\small
\begin {eqnarray*}{\cal B}[\chi_{c0}\to f_0(1370)f_0(1370)] {\cal B}
[f_0(1370)\to\pi^+\pi^-] {\cal B} [f_01370)\to K^+K^-] < 2.9\times 10^{-4},\\
{\cal B}[\chi_{c0}\to f_0(1370)f_0(1500)] {\cal B}
[f_0(1370)\to\pi^+\pi] {\cal B} [f_0(1500)\to K^+K^-] < 1.8\times 10^{-4},\\
{\cal B}[\chi_{c0}\to f_0(1500)f_0(1370)] {\cal B}
[f_0(1500)\to\pi^+\pi^-] {\cal B} [f_01370)\to K^+K^-] < 1.4\times 10^{-4},\\
{\cal B}[\chi_{c0}\to f_0(1500)f_0(1500)] {\cal B}
[f_0(1500)\to\pi^+\pi^-] {\cal B} [f_0(1500)\to K^+K^-] < 0.55\times 10^{-4},\\
{\cal B}[\chi_{c0}\to f_0(1500)f_0(1710)] {\cal B}
[f_0(1500)\to\pi^+\pi^-] {\cal B} [f_0(1710)\to K^+K^-] <
0.73\times 10^{-4}.
\end {eqnarray*}}

\begin{table}[htp]
\begin{tabular}{cccc}\hline
  \tablehead{1}{c}{b}{Decay mode}
  & \tablehead{1}{c}{b}{{$N^{fit}$}}
     & \tablehead{1}{c}{b}{Branching ratio $(10^{-4})$}
        & \tablehead{1}{c}{b}{s.s.} \\
  (X)&  & ${\cal B}
  [\chi_{c0}\ \rightarrow X \rightarrow \pi^+\pi^- K^+K^-]$ & \\\hline
${f_0(980)f_0(980)}$&$27.9\pm8.7$&$3.46\pm 1.08^{+1.93}_{-1.57}$&$5.3\sigma$\\
${f_0(980)f_0(2200)}$&$77.1\pm13.0$&$8.42\pm1.42^{+1.65}_{-2.29}$&$7.1\sigma$ \\
${f_0(1370)f_0(1710)}$&$60.6\pm15.7$&$7.12\pm1.85^{+3.28}_{-1.68}$&$6.5\sigma$\\
${K^*(892)^0\bar K^*(892)^0}$&$64.5\pm13.5$&$8.09\pm1.69^{+2.29}_{-1.99}$&$7.1\sigma$\\
${K^*_0(1430)\bar K^*_0(1430)}$&$82.9\pm18.8$&$10.44\pm2.37^{+3.05}_{-1.90}$&$7.2\sigma$\\
${K^*_0(1430)\bar K^*_2(1430)} +
c.c.$&$62.0\pm12.1$&$8.49\pm1.66^{+1.32}_{-1.99}$&$8.7\sigma$\\
${K_1(1270)^{+}K^{-} + c.c.,}$&&&\\
~~~$K_1(1270)\to K\rho(770)$&$68.3\pm13.4$&$9.32\pm1.83^{+1.81}_{-1.64}$&$8.6\sigma$\\
${K_1(1400)^{+}K^{-} + c.c.,}$&&&\\
~~~$K_1(1400)\to K^*(892)\pi$&$19.7\pm8.9$&$< 11.9$ (90\% C.L.) &$2.7\sigma$\\
\hline
\end {tabular}
\caption {Summary of the $\chicz\to \ppkk$ results, where $X$
represents the intermediate decay modes, $N^{fit}$ is the number
of fitted events, and s.s. indicates signal significance.}
\label{pwachi}
\end {table}

The above results supply important information on the
understanding of the natures of the scalar states~\cite{zhaoq}.

\subsection{Comments on the PWA}

PWA is extensively used in extracting physics information from the
experimental data, all the information listed above in this
section is from PWA. While we know the PWA uses the information in
the full phase space for physics study so that it is more powerful
than working in one dimension (invariant mass, for example) or a
bit higher dimension, it suffers from too many free parameters and
other technical problems.

First of all, the experimental data are all affected by the finite
resolution in momentum, energy, or direction measurement, this was
not considered in current analyses; secondly, the parametrization
of the resonance, especially the wide resonance, still have room
to improve. Finally, the effect of the imperfect simulation of the
detectors is hard to be considered in a fit with a few ten or even
more free parameters. These effects may not be significant when
the statistics is low, however, as the statistics increases, all
these effects will possibly produce significant fake signals. How
to handle these, if not now, at least in the near future, should
be studied since high luminosity experiments will be soon
available.

Studying the papers dealing with the PWA, it is found that two
important information are missing in most of the analyses, which
are the goodness-of-fit and the correlation coefficients between
the fit out parameters.

In many of the analyses, the comparison to the alternative fits
are given to show the fit is the best among all the fits tested,
however this does not guarantee the fit used in the analysis is
reliable, just like a fit to a Gaussian using a 2nd order
polynomial is better than using a straight line, but does not mean
the fit is acceptable. To give the goodness-of-fit is not easy in
case of PWA since the fitting function is always very complicated.
A possible way is to supply a simple $\chi^2$ test in one or
multi-dimension as has been done in Ref.~\cite{ppkk}, although not
perfect, it shows the reader a feeling how the fit describes the
data. Another possible way is try to use toy Monte Carlo to get
the expected distribution of the $\chi^2$ or the likelihood after
generating many MC samples using the fit out parameters as input,
and compare with the one got from the fit to the data --- this may
be a bit time consuming, however, it is more reliable since the
detector effects are considered.

The parameters from the fit are generally correlated, and
sometimes, some variables are highly correlated, in this case,
only reporting the diagonal error is not enough. This is extremely
important when there are two nearby resonances with strong
interference, they are anti-correlated and one component will
increase (decrease) as the other decrease (increase) to make the
total contribution unchanged. Neglecting this in the theoretical
analysis will also be dangerous. Another effect of the correlation
is the significance of the signal may be affected strongly, that
is, a declared significance may not be as high as that determined
when there is no correlation, and vice versa.

\section{``$\rho\pi$ puzzle'' in vector charmonia decay}

From perturbative QCD (pQCD), it is expected that both $\jpsi$ and
$\psp$ decaying into light hadrons are dominated by the
annihilation of $c\bar{c}$ into three gluons or one virtual
photon, with a width proportional to the square of the wave
function at the origin~\cite{appelquist}. This yields the pQCD
``12\% rule'',
\begin{equation}
 Q_h =\frac{{\cal B}_{\pspto h}}{{\cal
B}_{\jpsito h}} =\frac{{\cal B}_{\pspto \EE}}{{\cal B}_{\jpsito
\EE}} \approx 12\%.
\end{equation}
A large violation of this rule was first observed in decays to
$\rhopi$ and $K^{*+}K^-+c.c.$ by Mark~II~\cite{mk2}, known as {\it
the $\rhopi$ puzzle}, where only upper limits on the branching
fractions were reported in $\psp$ decays. Since then, many
two-body decay modes of the $\psp$ have been measured by the BES
collaboration and recently by the CLEO collaboration; some decays
obey the rule while others violate it~\cite{besres,cleocvp}.

The extension of the above rule to $\pspp$ is straightforward, in
the same scheme, one would expect
\begin{equation}
 Q_h^\prime =\frac{{\cal B}_{\pspp\to h}}{{\cal
B}_{\jpsito h}} =\frac{{\cal B}_{\pspp\to \EE}}{{\cal B}_{\jpsito
\EE}} \approx 1.9\times 10^{-4}.
\end{equation}

As the $\pspp$ data samples are available both at BESII and CLEOc,
the search for the decays of $\pspp$ into light hadrons was
performed. Since $\pspp$ is above the charm threshold, it is
expected the dominant decays of it is to charmed meson pairs,
however, large fraction of charmless decays of $\pspp$ is expected
if $\pspp$ is a mixture of $S$- and $D$-wave charmonium states and
the mixing is responsible for the ``12\% rule'' violation in
$\jpsi$ and $\psp$ decays. The above two rules may be tested by
the large data samples of $\jpsi$, $\psp$, and $\pspp$ at both
BESII and CLEOc.

\subsection{$\psi\to \rho\pi$: current status}

The $\rho\pi$ mode of the vector charmonia decays is essential for
this study, since this is the first puzzling channel found in
$\jpsi$ and $\psp$ decays. The new measurements, together with the
old information, show us a new picture of the charmonium decay
dynamics.

\subsubsection{$\jpsi\to \ppp$}

BESII measured the $\jpsi\to \ppp$ branching fraction using its
$\jpsi$ and $\pspto \jpsi\pip\pim$ data samples~\cite{bes3pi}, and
BARBAR measured the same branching fraction using $\jpsi$ events
produced by initial state radiative (ISR) events at
$\sqrt{s}=10.58$~GeV~\cite{babar3pi}. Together with the
measurement from Mark-II~\cite{mk2}, we get a weighted average of
the $\BR(\jpsito\ppp)=(2.00\pm 0.09)\%$.

To extract the $\jpsito \rhopi$ branching fraction, PWA is needed
to consider the possible contribution from the excited $\rho$
states, the only available information on the fraction of $\rhopi$
in $\jpsito \ppp$ was got at Mark-III. Using the information given
in Ref.~\cite{bill}, we estimate $\frac{\BR(\jpsito
\rhopi)}{\BR(\jpsito\ppp)}=1.17(1\pm 10\%)$, with the error from
an educated guess based on the information in the paper since we
have no access to the covariant matrix from the fit showed in the
paper. From this number and the $\BR(\jpsito\ppp)$ got above, we
estimate $\BR(\jpsito\rhopi)=(2.34\pm 0.26)\%$. This is
substantially larger than the world average listed by
PDG~\cite{pdg}, which is $(1.27\pm 0.09)\%$.

\subsubsection{$\psp\to \ppp$}

\begin{figure*}[htbp]
\begin{minipage}{7cm}
 \resizebox{15pc}{!}{\includegraphics[height=.3\textheight]{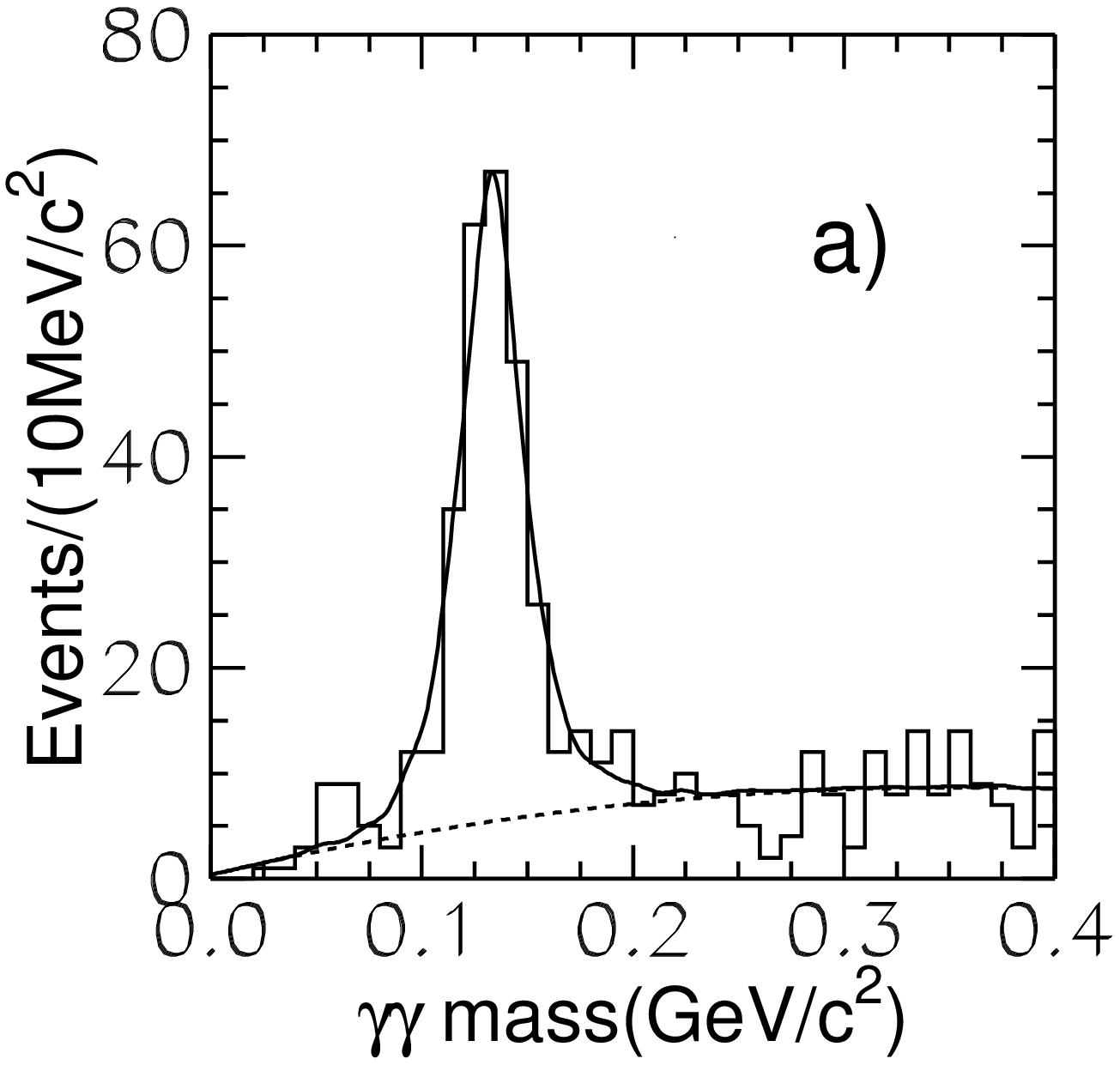}}
\end{minipage}
\begin{minipage}{7cm}
 \resizebox{14pc}{!}{\includegraphics[height=.3\textheight]{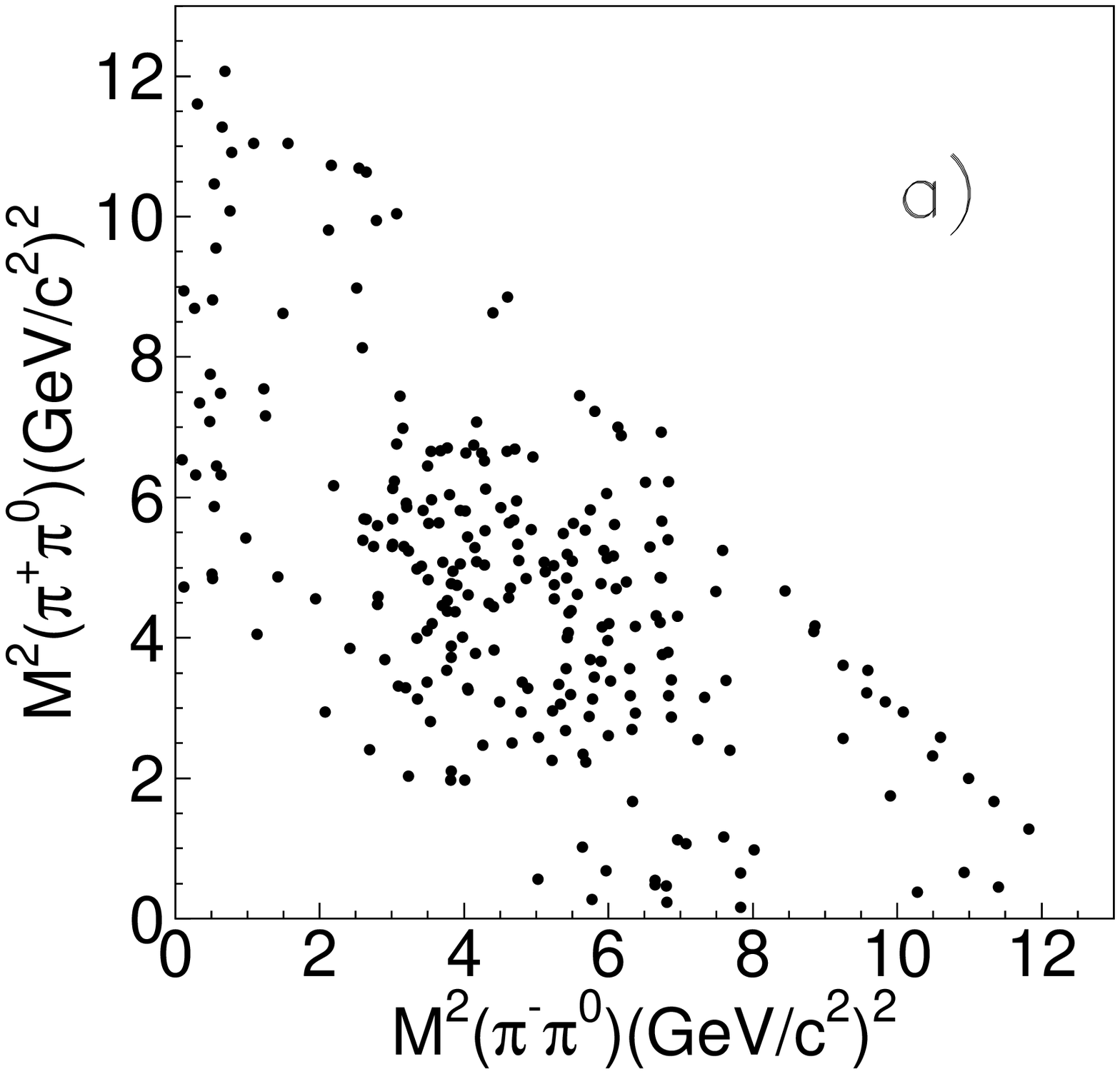}}
\end{minipage}
\caption{Two photon invariant mass distribution (left) and the
Dalitz plot (right) after final selection for BESII $\psp$ data.
The histograms are data, and the curves show the best fits.}
\label{psprhopi}
\end{figure*}

$\pspto \rhopi$ was studied both at BESII~\cite{psprhopi} and
CLEOc~\cite{cleocvp}. After selecting two charged pions and two
photons, clear $\piz$ signals are observed in the two photon
invariant mass spectra, the numbers of signals are found to be 229
and 196 from BESII (shown in left plot of Fig.~\ref{psprhopi} as
an example) and CLEOc respectively, and the branching fraction of
$\pspto \ppp$ is measured to be $(18.1\pm 1.8\pm 1.9)\times
10^{-5}$ and $(18.8^{+1.6}_{-1.5}\pm 1.9)\times 10^{-5}$ at BESII
and CLEOc respectively. The two experiments give results in good
agreement with each other. The Dalitz plot of $\pspto \ppp$ events
(right plot of Fig.~\ref{psprhopi}) shows very different signature
from that of $\jpsito \ppp$ decays, there are lots of events in
the center part of the Dalitz plot in the former case, while in
the latter, almost all the events are in the $\rho$ mass region
and there is almost nothing in the center of the Dalitz plot.

To extract the branching fraction of $\pspto \rhopi$, however,
BESII uses a PWA including the high mass $\rho$ states and the
interference between them, while CLEOc counts the number of events
by applying a $\rho$ mass cut. The branching fraction from BESII
is $(5.1\pm 0.7\pm 1.1)\times 10^{-5}$, while that from CLEOc is
$(2.4^{+0.8}_{-0.7}\pm 0.2)\times 10^{-5}$, the difference is
large. Although big difference exist between BESII and CLEOc
results exist, these do mean the $\pspto \rhopi$ signal exists,
rather than the signal is completely missing as has been guessed
before. If we take a weighted average neglecting the difference
between the two measurements, we get \( \BR(\pspto \rhopi)=(3.1\pm
0.7)\times 10^{-5} \).

Comparing $\BR(\pspto \rhopi)$ with $\BR(\jpsito \rhopi)$, one
gets
\[ Q_{\rhopi} = \frac{\BR(\pspto \rhopi)}{\BR(\jpsito \rhopi)} =
(0.13\pm 0.03)\%. \] The suppression compared to the 12\% rule is
obvious.

\subsubsection{$\pspp\to \ppp$}

It has been pointed out that the continuum amplitude plays an
important role in measuring $\pspp$ decays into light
hadrons~\cite{wympspp}. In fact, there are two issues need to be
clarified in $\pspp$ decays, that is whether $\pspp$ decays into
light hadrons really exist, and if it exists, how large is it. By
comparing the cross sections of $\EE\to \ppp$ at the $\pspp$
resonance peak ($\sqrt{s}=3.773$~GeV) and at a continuum energy
point ($\sqrt{s}=3.65$~GeV at BESII and $3.671$~GeV at CLEOc)
below the $\psp$ peak, both BESII and CLEOc found that
$\sigma(\EE\to \ppp)$ at continuum is larger than that at $\pspp$
resonance peak. The average of the two
experiments~\cite{psppbes,psppcleoc} are
    \[ \sigma(\EE\to \ppp)_{\hbox{on}}  =  7.5\pm 1.2~\hbox{pb}, \]
    \[ \sigma(\EE\to \ppp)_{\hbox{off}} = 13.7\pm 2.6~\hbox{pb}. \]
The difference, after considering the form factor variation
between 3.65 and 3.773~GeV, is still significant, and it indicates
there is an amplitude from $\pspp$ decays which interferes
destructively with the continuum amplitude, which makes the cross
section at the $\pspp$ peak smaller than the pure contribution of
continuum process.

For the $\rhopi$ mode, BESII can only give upper limit of its
cross section due to the limited statistics of the data sample,
the upper limit at 90\% C. L. is found to be 6.0~pb~\cite{psppbes}
at the $\pspp$ peak, which is in consistent with the measurement
from CLEOc using a much larger data sample: \(\sigma(\EE\to
\rhopi)_{\hbox{on}}  = 4.4\pm 0.6~\hbox{pb} \)~\cite{psppcleoc};
while the cross section at the continuum is $8.0^{+1.7}_{-1.4} \pm
0.9$~pb measured by CLEOc.

\begin{figure*}[htbp]
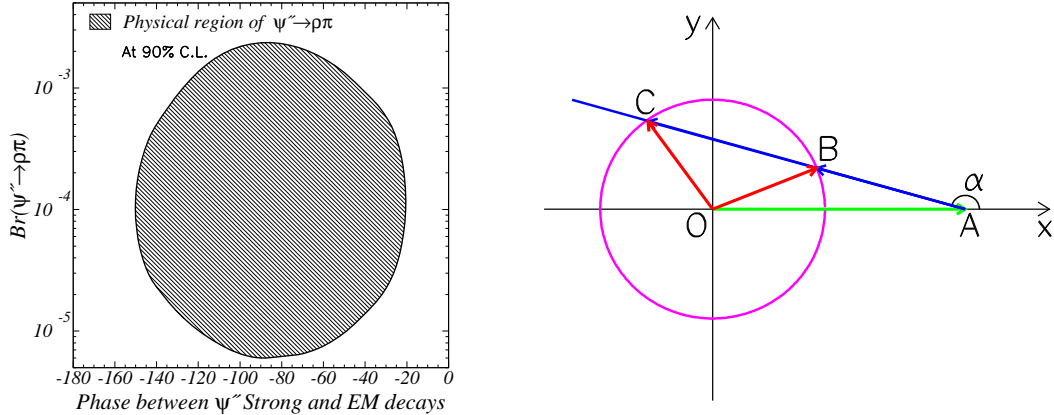

\begin{minipage}{7cm}
 \resizebox{14pc}{!}{\includegraphics[height=.5\textheight]{ang_cro2.epsi}}
\end{minipage}
\begin{minipage}{7cm}
 \resizebox{16pc}{!}{\includegraphics[height=.5\textheight]{two_sol.epsi}}
\end{minipage}
\caption{Physics region on $\BR(\psppto \rhopi)$ and the relative
phase between $\pspp$ strong and electromagnetic decays from BESII
(left); and the illustration of the two solutions in $\pspp$
decays (right): $\vec{OA}$ represents the continuum amplitudes,
$\vec{OB}$ or $\vec{OC}$ represents the peak amplitudes, and
$\vec{AB}$ and $\vec{AC}$ are the two solutions for the resonance
decay amplitudes, $\alpha$ is the relative phase between continuum
amplitude and the $\pspp$ strong decay amplitude, including the
relative phase $\phi$ and the relative phase between continuum and
$\pspp$ Breit-Wigner amplitudes.} \label{besii_pspp}
\end{figure*}

To extract the information on the $\psppto \rhopi$ branching
fraction, BESII developed a method based on the measured cross
sections at $\pspp$ resonance peak and at the
continuum~\cite{psppbes}. By neglecting the electromagnetic decay
amplitude of $\pspp$, there are two amplitudes contribute to the
cross section at the $\pspp$ peak, the strong decay amplitude of
$\pspp$ and the continuum amplitude. Taking the continuum
amplitude as a real number, the $\pspp$ strong decay amplitude is
described as one real number for the magnitude, and one phase
between $\pspp$ strong and electromagnetic decays to describe the
relative phase between the two amplitudes. Since only two
measurements are available (at $\pspp$ peak and at continuum), one
can only extract $\pspp$ decay branching fraction as a function of
the relative phase. BESII measurement on the upper limit of the
$\EE\to\rhopi$ cross section at $\pspp$ peak, together with the
CLEOc measurement of the continuum cross section restrict the
physics region of the branching fraction and the relative phase as
shown in Fig.~\ref{besii_pspp}(left). From the Figure, we see that
the branching fraction of $\psppto \rhopi$ is restricted within
$6\times 10^{-6}$ and $2.4\times 10^{-3}$, and the phase is
between $-150^\circ$ and $-20^\circ$, at 90\% C.L.

The observation of the $\EE\to\rhopi$ signal at $\pspp$ peak and
the measurement of the cross section~\cite{psppcleoc} at CLEOc
further make the physical region in the branching ratio and
relative phase plane smaller: the CLEOc measurement gives a
similar out bound of the physical region as BES gives, while also
indicates the central part of the physical region in
Fig.~\ref{besii_pspp}(left) is not allowed by physics. By using a
toy Monte Carlo to simulate the CLEOc selection criteria and the
interference between resonance and continuum amplitudes, we found
that the $\psppto \rhopi$ branching fraction could be either
$(2.1\pm 0.3)\times 10^{-3}$ or $(2.4^{+3.4}_{-2.0})\times
10^{-5}$ from the CLEOc measurements, if the relative phase
between $\pspp$ strong and electromagnetic decay amplitudes is
$-90^\circ$ as observed in $\jpsi$ and $\psp$
decays~\cite{wymphase}.

The reason why there are two solutions for $\pspp$ decays can be
understood as illustrated in Fig.~\ref{besii_pspp}b. If we take
the continuum amplitude as a real number, it can be shown as a
vector along the real axis in the complex plane, the total cross
section at $\pspp$ peak only gives the magnitude of the total
amplitude and it is shown as a circle in the plot, the amplitude
of the $\pspp$ decays, represented by a vector connecting the end
of the continuum amplitude and the end of the total amplitude may
have two cross points with the circle, representing the two
solutions of the $\pspp$ decay amplitudes, and thus the branching
fractions. Only in some very special cases, there is only one
solution. The angle between the continuum amplitude and the
$\pspp$ decay amplitudes is shown in the plot as $\alpha$, it is
the sum of the relative phase $\phi$ between the strong and
electromagnetic decays of $\pspp$, and the phase from Breit-Wigner
function for the $\pspp$ resonance. It can be seen that, the two
solution only happens when the cross section at continuum is
larger than or equal to that at $\pspp$ peak; otherwise, there is
only one solution, as in $\pspto \rhopi$ case. However, as in
physics, there is only one $\pspp$ decay branching fraction, there
must be a way to extract the $\pspp$ decay branching fraction
experimentally, this could be made possible if one more data
sample is taken at a different energy point, for example, in the
shoulder of the $\pspp$ resonance; furthermore, if one even wants
to get the value of the relative phase, one more data point is
necessary. In total, to get concrete physics information on the
$\pspp$ decay branching fraction, at least data samples at four
different energy points are needed, better have one point far from
the resonance, so that it gives absolute measurement of the
continuum amplitude.

Based on current data samples, one gets
\[ Q_{\rhopi}^\prime = \frac{\BR(\psppto \rhopi)}{\BR(\jpsito \rhopi)} =
(9.0\pm 1.6)\%\hbox{\, or\, } (0.10^{+0.15}_{-0.09})\%, \] to be
compared with the pQCD prediction of 0.019\% if $\pspp$ is pure
$D$-wave charmonium.

\subsection{Other studies and comments}

There are many more new measurements on $\psp$ decays for the
extensive study of the ``12\%
rule''~\cite{besres,bescleoc,cleocvp}, among which the
Vector-Pseudoscalar (VP) modes are measured at first priority. The
ratios of the branching fractions are all suppressed for these VP
modes compared with the 12\% rule. The multi-hadron modes and the
Baryon-antibaryon modes are either suppressed, or enhanced, or
normal, which are very hard to categorize simply. The various
models, developed for interpreting specific mode may hard to find
solution for all these newly observed modes.

The $\pspp$ decays into light hadrons were searched for in various
$\pspp$ decay modes, including VP and multi-hadron
modes~\cite{psppcleoc,psppmulti}. However, only the comparison
between the cross sections at continuum and those at $\pspp$
resonance peak are given, instead of giving the $\pspp$ decay
branching fractions. In current circumstances, it is still not
clear whether the $\pspp$ decays into light hadrons with large
branching fractions, since, as has been shown in the $\rhopi$
case, there could be two solutions for the branching fraction, and
the two values could be very different.

As the luminosity at the $\pspp$ peak is large enough, current
study is limited by the low statistics at the continuum: at CLEOc,
the luminosity at continuum is more than an order of magnitude
smaller than that at peak, this prevents from a high precision
comparison between the cross sections at the two energy points.
One conclusion we can draw from the existing data is that the
measurements do not contradict with the assumption that the
relative phase between $\pspp$ strong and electromagnetic decay
amplitudes is around $-90^\circ$, and the $\pspp$ decays into
light hadrons could be large.

The study of the $\rhopi$ puzzle between $\psp$ and $\jpsi$ decays
and the charmless decays of $\pspp$ should not be isolated as they
were since $\jpsi$, $\psp$ and $\pspp$ are all charmonium states
with very similar quantum numbers, and it is expected $\psp$ and
$\pspp$ are the mixtures of $2S$ and $1D$ states~\cite{rosnersd}.
In developing models to solve one of the problems, the others
should also be considered. There have been a few models developed
following this line or can be easily extended to all these three
states, like the $S$- and $D$-wave charmonia mixing
model~\cite{rosnersd}, the $D\bar{D}$ re-annihilation in
$\pspp$~\cite{rosnerddb}, the four-quark component in
$\pspp$~\cite{voloshin}, and survival $c\bar{c}$ in
$\psp$~\cite{survival}, and so on. Experimentally testable
predictions are welcome for validating the models.

One further observation is that many of the attempts to interpret
the $\rho\pi$ puzzle are based on the potential models for the
charmonium which were developed more than 20 years ago, as the
B-factories discovered many new charmonium states~\cite{xyz} which
are hard to be explained in the potential models, it may indicate
even the properties of $\jpsi$, $\psp$ and $\pspp$ are not as
expected from the potential models. The further understanding of
the other high mass charmonium states may shed light on the
understanding of the low lying ones.

\section{Summary}

There are many new results on hadron spectroscopy from the
charmonium decays from BES and CLEOc experiments. While many
analyses supply more information on the known states like the
light scalar particles to understand the nature of them, there are
also new observations which may indicate there are still something
unexpected in the low energy regime such as the possible existing
baryonium. The decay properties of the vector charmonium states,
although have been studied for more than three decades, is still
far from being understood, one extreme example is the ``$\rhopi$
puzzle'' in $\jpsi$ and $\psp$ decays. Further studies of all
these are expected from the BESIII at BEPCII which will start its
data taking in 2007.


\begin{theacknowledgments}
The author would like thank the organizer for the invitation to
give the talk, and for the successful organization of the
conference. I would also like to thank Dr. Xiaohu Mo, Dr. Ping
Wang and Dr. Zheng Wang for their contributions and for many
helpful discussions. I also thank Dr. Liaoyuan Dong, Dr. Zijin
Guo, Dr. Yingchun Zhu and Dr. R.~Briere for their help in
preparing the talk and this paper.
\end{theacknowledgments}


\begin{thebibliography}{99}

\bibitem{bes} BES Collaboration, J.~Z.~Bai {\em et al.}, Nucl. Instr. Meth.
              A {\bf 344}, 319 (1994); A {\bf 458}, 627 (2001).
\bibitem{cleo} CLEO-c Collaboration, CLEO-c and CESR-c: A New Frontier of
              Weak and Strong Interactions, CLNS 01/1742.
\bibitem{etap} BES Collaboration, M.~Ablikim {\em  et al.},
              hep-ex/0508025, accepted by Phys. Rev. Lett.
\bibitem{gpp} BES Collaboration, J.Z. Bai {\em  et al.},
              Phys. Rev. Lett. {\bf 91}, 022001 (2003).
\bibitem{pdg} Particle Data Group, S.~Eidelman {\em et al.},
              Phys. Lett. B {\bf 592}, 1 (2004).
\bibitem{BESI} BES Collaboration, J.Z. Bai {\sl  et al.},
               Phys. Lett. {\bf B446}, 356 (1999).
\bibitem{fsi1} B.S. Zou and H.C. Chiang, Phys. Rev. D {\bf 69}, 034004 (2003).
\bibitem{fsi2} A. Sibirtsev {\em et al.}, Phys. Rev. D {\bf 71}, 054010 (2005).
\bibitem{yan} G.J. Ding and M.L. Yan, Phys. Rev. C {\bf 72}, 015208 (2005).
\bibitem{zhu} S.L. Zhu and C.S. Gao, hep-ph/0507050.

\bibitem{wpipi} BES Collaboration, M.~Ablikim {\em  et al.},
              Phys. Lett. B {\bf 598}, 149 (2004).
\bibitem{wkk} BES Collaboration, M.~Ablikim {\em  et al.},
              Phys. Lett. B {\bf 603}, 138 (2004).
\bibitem{phipipi} BES Collaboration, M.~Ablikim {\em  et al.},
              Phys. Lett. B {\bf 607}, 243 (2004).
\bibitem{gkk} BES Collaboration, J.~Z.~Bai {\em et al.},
              Phys. Rev. D {\bf 68}, 052003 (2003).
\bibitem{zhuys} Y.S. Zhu (for the BES Collaboration), this
               conference.
\bibitem{oset} L.~Roca {\em  et al.},
              Nucl. Phys. A {\bf 744}, 127 (2004).
\bibitem{bugg} D.~Bugg, this conference.
\bibitem{close} F.~Close and Q.~Zhao, hep-ph/0504043.
\bibitem{zhao} Q.~Zhao, B.S.~Zou and Z.B.~Ma, hep-ph/0508088.
\bibitem{ppkk} BES Collaboration, M.~Ablikim, {\em et al.},
              hep-ex/0508050.
\bibitem{zhaoq} Q.~Zhao, hep-ph/0508086.

\bibitem{appelquist}T.~Appelquist and H.~D.~Politzer,
             Phys. Rev. Lett. {\bf 34}, 43 (1975); A.~De R\'{u}jula and
             S.~L.~Glashow, Phys. Rev. Lett. {\bf 34}, 46~(1975).
\bibitem{mk2}Mark~II Collaboration, M.~E.~B.~Franklin {\em et al.},
             Phys. Rev. Lett. {\bf 51}, 963 (1983).
\bibitem{besres} Many results may be found in Ref.~\cite{pdg};
        more recent results may be found in BES Collaboration, J.~Z.~Bai {\em et al.},
        Phys. Rev. D {\bf 69}, 072001 (2004); Phys. Rev. Lett. {\bf 92}, 052001 (2004);
        BES Collaboration, M.~Ablikim {\em et al.}, Phys. Rev. D {\bf 70},
        112007 (2004); Phys. Rev. D {\bf 70}, 112003 (2004);
        and Phys. Lett. B {\bf 614}, 37 (2005).
\bibitem{cleocvp} CLEO Collaboration, N.~E.~Adam {\em et al.},
             Phys. Rev. Lett. {\bf 94}, 012005 (2005).
\bibitem{bes3pi} BES Collaboration, J.~Z.~Bai {\em et al.},
          Phys. Rev. D {\bf 70}, 012005 (2004).
\bibitem{babar3pi} BES Collaboration, J.~Z.~Bai {\em et al.},
          Phys. Rev. D {\bf 70}, 072004 (2004).
\bibitem{bill} MARK-III Collaboration, L.~Chen and W.M.~Dunwoodie,
          in {\em Proceedings of Hadron 91 Conference, College Park,
          Maryland, 1991}, p. 100, SLAC-PUB-5674 (1991).
\bibitem{psprhopi} BES Collaboration, M.~Ablikim {\em  et al.},
              Phys. Lett. B {\bf 619}, 247 (2005).
\bibitem{wympspp} P.~Wang, X.~H.~Mo and C.~Z.~Yuan,
         Phys. Lett. B {\bf 574}, 41 (2003).
\bibitem{psppbes} BES Collaboration, M.~Ablikim {\em  et al.},
                  hep-ex/0507092, Phys. Rev. D (in press).
\bibitem{psppcleoc} CLEO Collaboration, G.S.~Adams {\em  et al.},
                  hep-ex/0509011.
\bibitem{wymphase} P.~Wang, C.~Z.~Yuan and X.~H.~Mo,
         Phys. Rev. D {\bf 69}, 057502 (2004).
\bibitem{bescleoc} BES Collaboration, M.~Ablikim {\em  et al.},
         Phys. Rev. D {\bf 71}, 072006 (2005); CLEO Collaboration,
         R.~A.~Briere {\em et al.},
         Phys. Rev. Lett. {\bf 95}, 062001 (2005);
         CLEO Collaboration, T.~K.~Pedlar {\em et al.},
         Phys. Rev. D {\bf 72}, 051108(R) (2005).
         X.~H.~Mo (for the BES Collaboration), this conference.
\bibitem{psppmulti} CLEO Collaboration, G.~S.~Huang {\em et al.},
         hep-ex/0509046.
\bibitem{rosnersd} J.~L.~Rosner, Phys. Rev. D {\bf 64},
        094002 (2001).
\bibitem{rosnerddb} J.~L.~Rosner, hep-ph/0405196.
\bibitem{voloshin} M.~Voloshin, Phys. Rev. D {\bf 71},
                   114003 (2005).
\bibitem{survival} P.~Artoisenet, J.~M.~Gerard and J.~Weyers,
        hep-ph/0506325; J.~M.~G\'{e}rard and J.~Weyers, Phys.
        Lett. B {\bf 462}, 324 (1999).
\bibitem{xyz} K.~K.~Seth, this conference.

\end{thebibliography}
\end{document}